\documentclass[pss]{wiley2sp} 
\usepackage{amsmath}
\usepackage{bm}
\usepackage{appendix}

\begin{document}


\title{In memoriam Leonid V. Keldysh}

\author{%
  Michael Bonitz\textsuperscript{\Ast,\textsf{\bfseries 1}},
  Antti-Pekka Jauho\textsuperscript{\textsf{\bfseries 2}},
  Michael Sadovski\textsuperscript{\textsf{\bfseries 3}}, and
  Sergei Tikhodeev\textsuperscript{\textsf{\bfseries 4}}
  }

\mail{e-mail
  \textsf{bonitz@theo-physik.uni-kiel.de}, Phone: +49 431 880 4122}

\institute{%
  \textsuperscript{1}\,Institut f\"ur Theoretische Physik und Astrophysik, Christian-Albrechts-Universit\"at zu Kiel, 24098 Kiel, Germany\\
  \textsuperscript{2}\,Department of Micro- and Nanotechnology, Technical University of Denmark
  \\
  \textsuperscript{3}\,Institute for Electrophysics, RAS Ural Branch,
Ekaterinburg 620016, Russia and M.N. Mikheev Institute for Metal Physics, RAS Ural Branch, Ekaterinburg 620108, Russia
  \\
  \textsuperscript{4}\,Department of Physics,
   M.V. Lomonosov Moscow State University, 119991 Moscow
   and
   A.M. Prokhorov General Physics Institute,
   Russian Academy of Sciences, 
   Vavilova Street, 38
   Moscow 119991, Russia
  }


\abstract{Leonid Keldysh -- one of the most influential theoretical physicists of the 20th century -- passed away in November 2016. Keldysh is best known for the diagrammatic formulation of real-time (nonequilibrium) Green functions theory and for the theory of strong field ionization of atoms. Both theories profoundly changed large areas of theoretical physics and stimulated important experiments.  Both these discoveries emerged almost simultaneously -- like Einstein, also Keldysh had his \textit{annus mirabilis} -- the year 1964.
But the list of his theoretical developments is much broader and is briefly reviewed here.}

\maketitle   

\begin{figure}[h]%
\centering
\includegraphics*[width=0.85\linewidth]{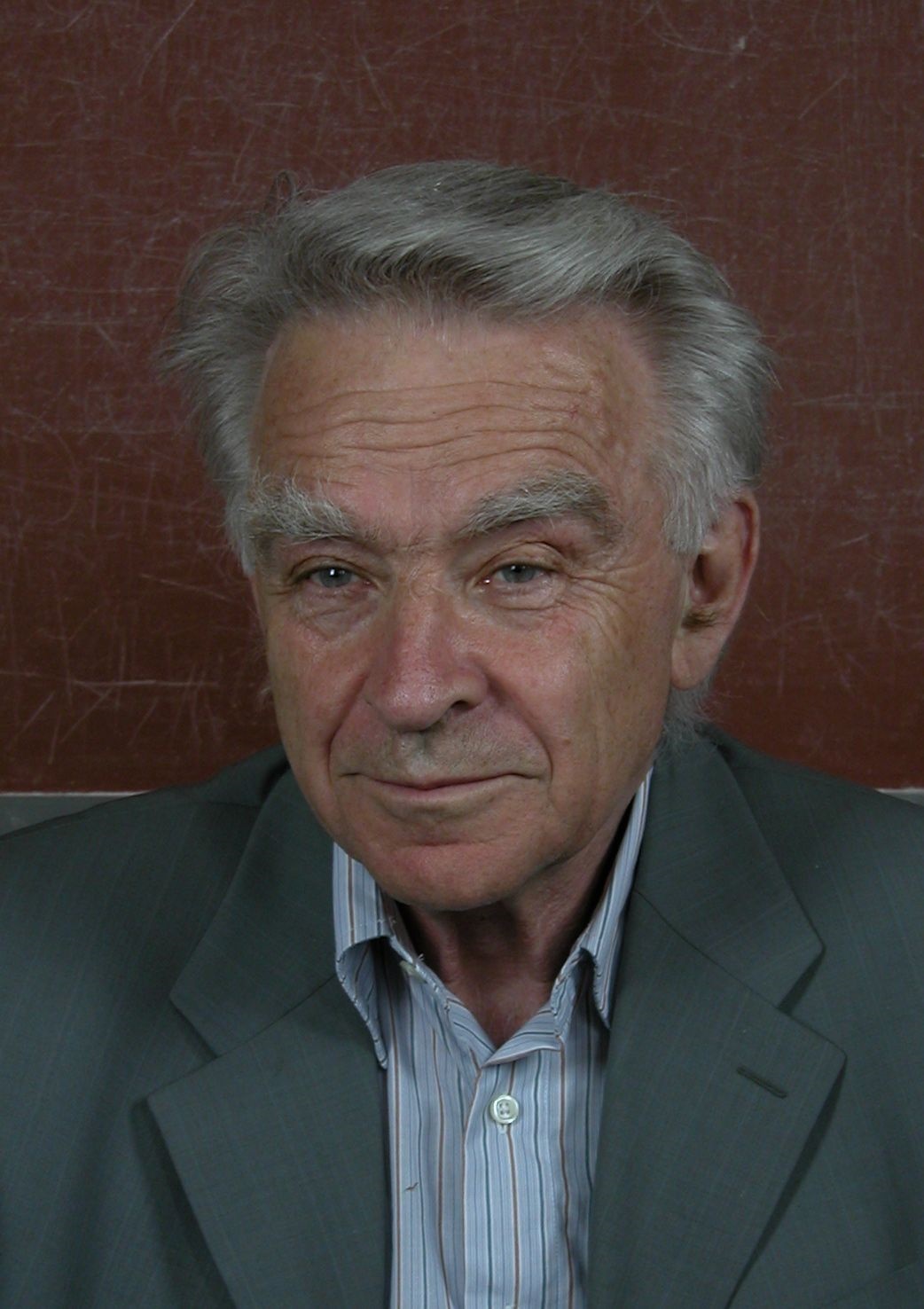}
\caption{%
Leonid V. Keldysh (1931-2016), photograph from around 2010, provided by M. Sadovskii.
}
\label{fig:main}
\end{figure}

\section{Introduction}\label{s:intro}
On November 11 2016 Leonid Veniaminovich Keldysh passed away in Moscow. Keldysh was a Russian theoretical physicist who had a tremendous influence on many fields of physics. In this article we briefly describe Keldysh's most important contributions and discuss how they influenced and continue to influence modern physics. In particular, we concentrate on his work on nonequilibrium many-particle physics that is related to his discovery of nonequilibrium Green functions theory, see Sec.~\ref{s:negf}. At the same time it is of high interest to recall many of his other activities, see Sec.~\ref{s:other}, including his contributions to strong field ionization, Sec.~\ref{ss:strong-field}, and exciton physics, cf. Sec.~\ref{ss:eh}.

Keldysh's heritage includes 
77 scientific publications 
\cite{ISI:A1958WT97700025,ISI:A1958WX85500028,KeldyshEffect,ISI:A1960WE64800018,ISI:A1960WN97500020,ISI:A1963WM20800048,keldysh_ptt_62,ISI:A1963WP01800029,keldysh_jetp_64,ISI:A1964WT61000005,ISI:A1964WT61500020,keldysh64,keldysh64_field,ISI:A19657071700023,keldysh_ptt_64,ISI:A19657168300010,ISI:A19668668700009,ISI:A19667931900016,ISI:A19679362600012,keldysh_jetp_68,ISI:A1969D031000004,ISI:A1969E591300010,ISI:A1969E811600006,ISI:A1970F063000017,ISI:A1970I354600011,ISI:A1972M332000004,keldysh_jetp_72,ISI:A1973P786000039,ISI:A1974T394700028,ISI:A1975AU26400005,ISI:A1975AT03500030,ISI:A1976CE06600002,ISI:A1976DB30000006,ISI:A1976DJ19200006,keldysh_jetp_76,ISI:A1976BR33200035,ISI:A1977CT30500023,ISI:A1977DV91300039,ISI:A1978FR24900010,ISI:A1979JN00600015,keldysh_1979_JETPL,ISI:A1980LJ42200006,ISI:A1980KN51300038,ISI:A1981MB17700017,ISI:A1982PJ79700021,ISI:A1983QB18700036,ISI:A1984AFY8900005,ISI:A1984TJ11900023,ISI:A1986C643700031,ISI:A1986D472300009,ISI:A1986F487200028,ISI:A1988P896100047,ISI:A1991FE66200024,ISI:A1992JQ13000097,ISI:A1992JR28400011,ISI:A1992JT05900011,ISI:A1993ML21500095,ISI:A1994QA79500021,ISI:A1995QQ83700001,ISI:A1997XR96400071,ISI:000071319400004,ISI:000072424800001,ISI:000087302200002,ISI:000179600900006,ISI:000184336600038,keldysh_pngf2,ISI:000225020200039,ISI:000230494900007,ISI:000230238900002,ISI:000237842200120,ISI:000237728200050,ISI:000239932300003,ISI:000243731300005,ISI:000242102500035,ISI:000392143400001,ISI:000424395100006,ISI:000424395100007},
that reflect the broad range of topics, Keldysh was interested in and how this interest evolved over time. His two most influential papers on  real time Green Functions \cite{keldysh64} and strong field ionization \cite{keldysh64_field} were published in 1964 in the Zhurnal Teoreticheskoi i Eksperimentalnoi Fiziki\footnote{abbreviated ZhETF, the English translation is being published as JETP or Soviet Physics JETP)} and collected about 3300 and 6000 citations, respectively\footnote{according to Google Schloar, December 2018}. In fact, these two papers were written almost at the same time. They were submitted by Keldysh to the journal on April 23 and May 23 1964, respectively, making the year 1964 Keldysh's \textit{annus mirabilis}. 
All articles of L.V. Keldysh are listed in chronological order in the reference section at the end of this paper.

Also, the activity of Keldysh in support of Russian science, in general, and the Academy of Sciences, in particular, is documented in 7 articles published between 1992 and 1999, \cite{ISI:A1992KH81500002,ISI:A1992JG61900007,ISI:A1992HD70800001,ISI:A1996TX10100004,ISI:A1996UX63100002,ISI:A1997XF16000005,ISI:000085196500001}. They are interesting historical documents in their own but also show that Keldysh was speaking up publicly when he thought this is necessary, often together with other leading Russian colleagues. 
Some information on the often difficult political environment is contained in the biographical notes in Sec.~\ref{s:bio}.
Moreover, Keldysh published a remarkable number of 61 short notes in honor of leading Russian physicists -- a special tradition in Soviet and Russian science. These articles include 42 birthday congratulations and 29 obituaries. 

Finally, we also include some remarks on Keldysh's students and Keldysh's work as a mentor, in Sec.~\ref{s:school}.

\section{Biographical notes \cite{keldysh_physics-today17}}\label{s:bio}
%
%
Leonid Keldysh was born in Moscow on 7 April
1931 in a family of scientists. His mother, Lyudmila
Vsevolodovna Keldysh, was a leading Soviet mathematician, her brother, an applied mathematician, Mstislav
Vsevolodovich Keldysh was one of the leaders of the
Soviet space program, later becoming the President of
the USSR Academy of Sciences. Leonids's step father was Petr Sergeevich Novikov, a full member of
the Academy and also a leading mathematician, while
Leonid's younger step brother, Sergei Petrovich Novikov,
also becoming a mathematician and Academy member,
was later awarded the Fields medal. But Leonid's choice
was theoretical physics.

In 1948 Keldysh enrolled in the Physics Department  of Moscow State Lomonosov University (MGU), where he graduated in 1954
(attending also courses at the Department of mechanics
and mathematics for an extra year). After this he started to work at the Theoretical Physics Department of the P.N. Lebedev Physics Institute (LPI) of the Academy of Sciences, which remained his work place until the
end of his life. His scientific supervisor at LPI was Vitaly Lazarevich Ginzburg, and the Theoretical
Department at that time was headed by Igor Evgenievich
Tamm (both later becoming Nobel prize winners).
However, since these early years, Leonid was essentially
a self--made man in science.


In his early works (1957--1958) Keldysh developed a consistent theory of phonon-assisted tunneling
in semiconductors which was immediately recognized by
the semiconductor community. His most famous work of this period was devoted to the calculation of the electric field-induced
shift of the absorption edge in semiconductors, what is now
called the Franz--Keldysh effect \cite{FranzEffect,KeldyshEffect}, see Sec.~\ref{sss:fke}. In the early 1960s he proposed to use spatial modulation of the lattice to create an artificial band structure \cite{keldysh_ptt_62}. This idea was later realized
in semiconductor superlattices. He also developed an
original theory of core levels in semiconductors \cite{keldysh_jetp_64}. One of
his most famous works of this period was the 1964 theory of tunnel and (multi-)photon ionization of atoms by intense electromagnetic waves \cite{keldysh64_field} that became the starting point for
the entire field of intense laser--matter interaction, including atoms, ions, molecules, plasmas and solids, cf. Sec.~\ref{ss:strong-field}. This field has recently been reviewed in Ref.~\cite{ReviewStrongFieldIonization}, where it is concluded that the success behind the theory is that it precisely fulfills the criterion ``making things as simple as possible, but not simpler''.  This feature is characteristic of many of Keldysh's other influential papers.

Leonid Keldysh started working in science during a period when quantum field theory methods were popular and successfully applied in condensed matter physics. Here he made his most famous contribution with his 1964 work on a general diagram technique for nonequlibrium processes \cite{keldysh64}. Introducing Green functions with time--ordering along what is today known as the (Schwinger)\-Keldysh time contour, he was able to construct the standard Feynman diagrams for these Green
functions at finite temperatures and for general nonequilibrium states, see Sec.~\ref{s:negf}.

\begin{figure}[h]%
\centering
\includegraphics*[width=0.75\linewidth]{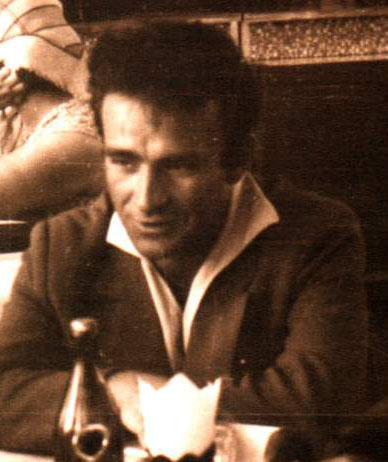}
\caption{%
Leonid V. Keldysh around 1962, photograph provided by M. Sadovskii.
}
\label{fig:62}
\end{figure}

Strangely enough, even at that time, ten years after
starting his work, he had not yet been awarded any higher scientific degree. However, when he finally submitted the Candidate of Science (PhD) thesis, in 1965, he was immediately
awarded the degree of Doctor of Science (similar to habilitation in Germany). In 1968 he was elected corresponding member of the USSR Academy of Sciences, becoming a full member in 1976.

Since 1964 Keldysh's interests moved to  semiconductors. In his work with Yu.V.
Kopaev \cite{keldysh_ptt_64} he introduced the new concept of an excitonic insulator and to laser excited nonequilibrium exciton systems,
 exciton superfluidity  \cite{keldysh_jetp_68} and their ioinization
into an electron--hole quantum liquid of electron--hole droplets, for details see Sec.~\ref{ss:eh}.
\begin{figure}[h]%
\centering
\includegraphics*[width=0.75\linewidth]{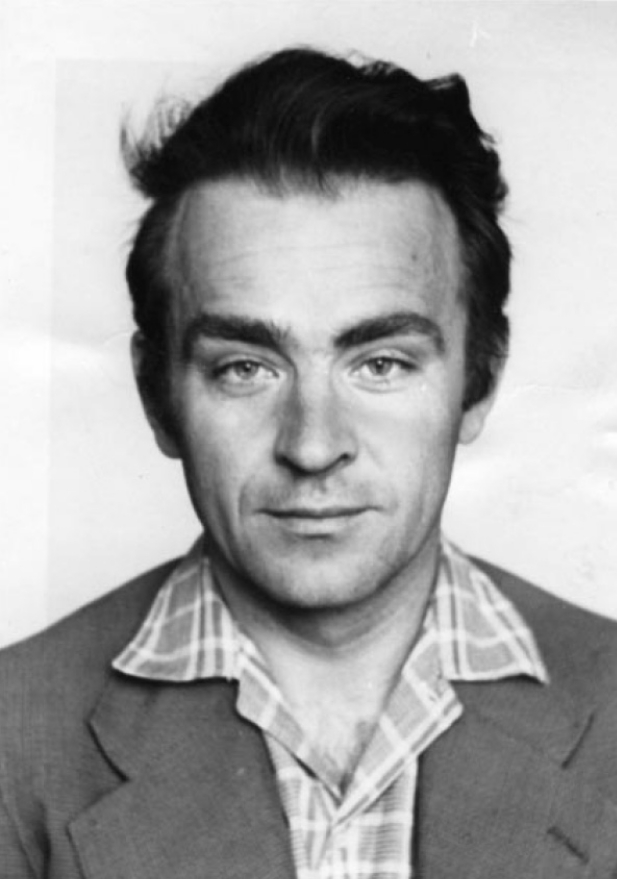}
\caption{%
Leonid V. Keldysh around 1968, photograph provided by M. Sadovskii.
}
\label{fig:68}
\end{figure}

Since 1965 Keldysh was a professor at MGU heading the chair of quantum radiophysics (1978-2001). He had many PhD students, a
number of which later became famous theoreticians, professors and members of the Russian Academy of Sciences, see Sec.~\ref{s:school}.
He was member of editorial boards of the leading Russian
physics journals and served as Editor in Chief of Physics Uspekhi, from 2009 to 2016.
Keldysh was awarded numerous prizes, including the
Lenin prize (1974), the Hewlett--Packard Prize (1975), the
Alexander von Humboldt Prize (1994), the Rusnanoprize
(2009), the Eugene Feinberg Memorial Medal (2011), the Pomeranchuk Prize (2014) and the Grand Lomonosov Gold
Medal of the Russian Academy of Sciences (2015). He
was elected foreign member of the US National
Academy of Sciences (1995) and became a Fellow of the American Physical Society in 1996.

In the late 1980s Keldysh had to perform various administrative duties, which he actually did not like at all,
but considered impossible to reject during this difficult period for  Russian Science. These included the head
of the Theoretical Physics Department and the director
of the Lebedev Institute (1989--1994) and also the position
of a Secretary of the General Physics Department of the Russian Academy of Sciences (1991--1996). During this period he
lived in his own way, never conforming to external circumstances. He always was a highly independent person, and it was
impossible to persuade him to take a decision  he did not agree with. He was among the leading RAS members
who strictly rejected the Government--proposed reform of
the Academy in 2013, becoming a member of the influential ``Club of July 1'' within the RAS, opposing this reform.
\begin{figure}[t]%
\centering
\includegraphics*[width=0.75\linewidth]{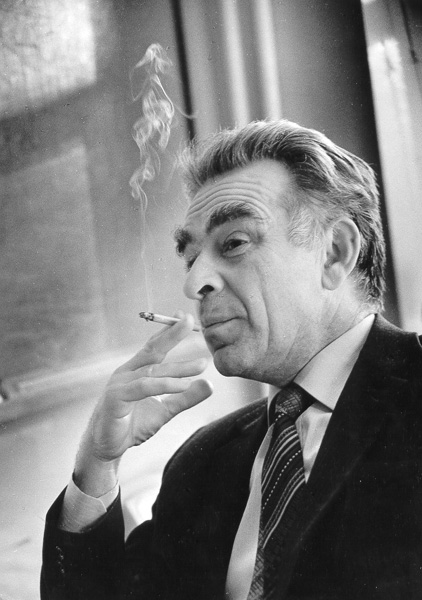}
\caption{%
Leonid V. Keldysh around 1989, photograph provided by M. Sadovskii.
}
\label{fig:89}
\end{figure}
%
%
\begin{figure}[h]%
\centering
\includegraphics*[width=0.95\linewidth]{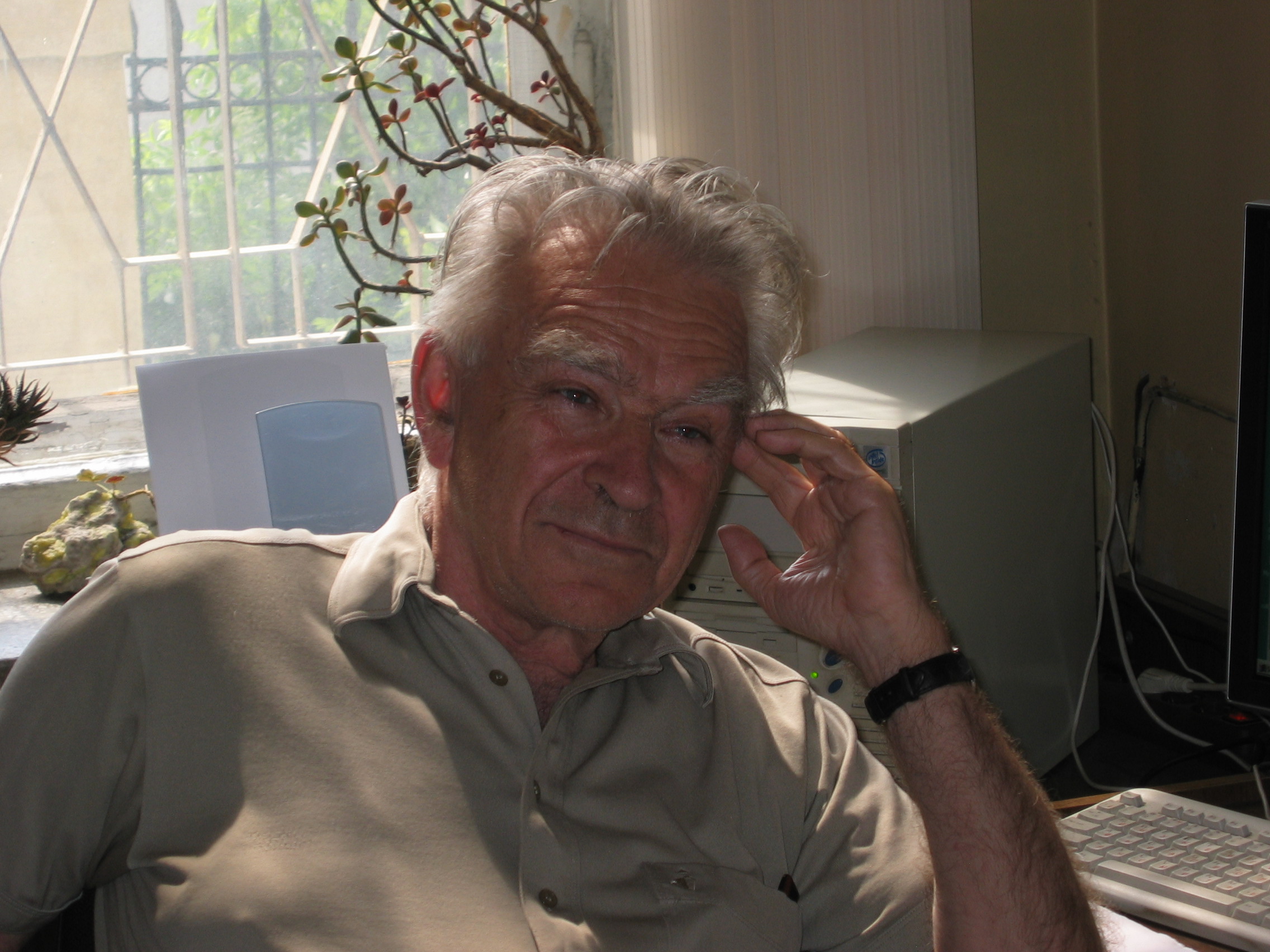}
\caption{%
Leonid V. Keldysh around 2006, photograph by Nikolay Gippius.
}
\label{fig:06}
\end{figure}

\section{Nonequilibrium (Real-time or Keldysh) Green Functions (NEGF)}\label{s:negf}
Judging by its impact on a huge number of fields Keldysh's Real-time Green functions theory is, without question, his most important discovery, and we address it in some more detail.
\subsection{The story of NEGF}
Quantum many-body systems have been described by many different approaches including wave function methods, reduced density operators (quantum BBGKY-hierarchy) of Bogolyubov, Krikwood and others, e.g. \cite{bonitz_qkt}  as well as Green functions and Feynman diagrams by Schwinger, Dyson and Feynman. Following the results for the ground state soon the extension to thermodynamic equilibrium was developed in the 1950s by Matsubara, Kubo as well as Abrikosov, Gorkov, Dzyaloshinski in the U.S.S.R. which led to the concept of imaginary-time Green functions. The idea to rewrite the canonical density operator in thermodynamic equilibrium as a quantum-mechanical evolution operator, but in imaginary time, was then quite popular in a number of fields, including Feynman's path integral concept, so that step was rather natural.

However, the extension of the technique from thermodynamic equilibrium to \textit{arbitrary nonequilibrium situations} is a huge step that is far less straightforward, and it took more time to develop. These developments occured almost independently in the U.S. and in the U.S.S.R. The works in the U.S. were mostly due to Martin and Schwinger who derived the generalization of the BBGKY-hierarchy to the case of many-time Green functions \cite{martin-schwinger_59}, and Baym and Kadanoff who derived and analyzed the generalization of the Boltzmann equation that includes memory effects \cite{kadanoff-baym-book}. These developments were reviewed in detail by Paul Martin and Gordon Baym in their lectures at the first Nonequilibrium Green Functions conference in Rostock, Germany, in 1999, cf.~Refs.~\cite{martin_pngf1,baym_pngf1}. The Russian developments in the field of Nonequilibrium Green functions are due solely to Leonid Keldysh and were published in his seminal paper \cite{keldysh64} where he introduced the  ``round trip time contour'' -- a small but ingeneous mathematical trick -- that allowed him to rigorously extend Feynman's diagram technique to nonequilibrium.
The Russian developments in thermodynamic and Nonequilibrium Green functions were reviewed by Alex Abrikosov  \cite{abrikosov_pngf2} and Leonid Keldysh \cite{keldysh_pngf2}, respectively. The latter article is reprinted as a supplement \cite{supplement} to this paper.

\subsection{The PNGF conferences and Leonid Keldysh}\label{ss:pngf}
Interestingly, after writing his paper introducing NEGF in 1964, Keldysh did not actively continue these developments (the same was true for Baym and Kadanoff). So it must have been a surprise for them that they were in 1999 invited to a conference entitled ``Kadanoff-Baym equations--Progress and Perspectives for Many-Body Theory'', 35 years after the original developments. In fact, in the 1970s and 1980s NEGF were used only by a few groups world wide but the activities increased significantly in the 1990s when NEGF methods were used in semiconductor optics and various groups learned to directly solve the Keldysh-Kadanoff-Baym equations (KBE) on modern computers, following the pioneering work of Danielewicz \cite{DANIELEWICZ_84_ap2} on nuclear collisions. Not surprisingly, many theorists\footnote{Here we should mention, in particular, W. Sch\"afer \cite{bonitz_pngf3,schaefer_wegener}, D. Kremp \cite{bornath_cpp17,kremp_schlanges_kraeft_book}, H. Haug \cite{haug_2008_quantum} and K. Henneberger in Germany and their schools.} expected that these equations would lead to breakthroughs in many fields which indeed turned out to be the case, see Sec.~\ref{ss:negf-current}.
\begin{figure*}[h]%
\centering
\includegraphics*[width=0.85\linewidth]{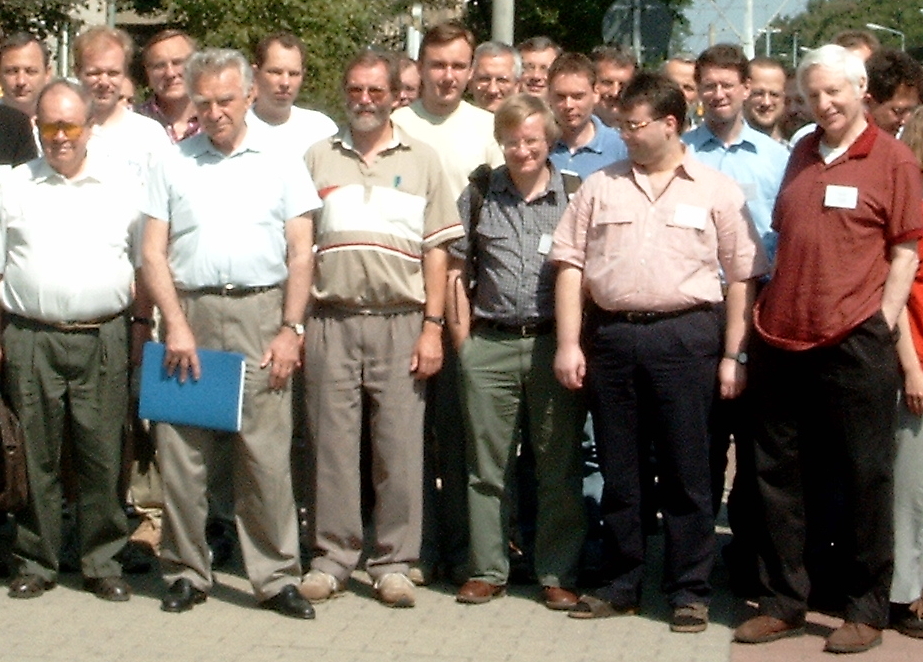}
\caption{%
Part of the participants of the Conference ``Progress in Nonequilibrium Green Functions II'', Dresden, August 2002.
Front row from left: Alexei Abrikosov, Leonid Keldysh, J\"orn Knoll, Pawel Danielewicz, Hendrik van Hees, Paul C. Martin. Part of the conference photo, from Ref.~\cite{pngf2}. Second and third row, among others: Paul Gartner, Egidius Anisimovas, Vladimir Filinov, Robert van Leeuwen, Alexey Filinov, Roland Zimmermann, Antti-Pekka Jauho and Rolf Binder. Photograph by M. Bonitz}
\label{fig:pngf2}
\end{figure*}

At the same time, the lengthy title of the conference in 1999 reflects some confusion in the community about the different contributions of the American and Russian founders of the theory and about the priorities. Even Baym was under the impression that Keldysh's work of 1964 was a follow up to their book \cite{kadanoff-baym-book}, as he pointed out in his conference talk in 1999 and in the proceedings \cite{baym_pngf1}. However, this was an incorrect assumption. Not surprisingly, Keldysh -- who could not participate in the 1999 conference -- was very upset when he became aware of Baym's article. He then took the opportunity to attend the second conference, ``Progress in Nonequilibrium Functions (PNGF) II'' in Dresden in 2002 and, in his lecture, to ``straighten'' things out. For everybody who uses NEGF today or will do so in the years to come, this turned out to be a very lucky case, because Keldysh summarized in some detail and in his honest style how his ideas emerged and who influenced him. We are lucky that he published his recollections in the conference book \cite{pngf2}, and his article 
is reprinted as a supplement to this paper. 
A photo showing Leonid Keldysh at the PNGF II together with, among others, Alex Abrikosov, Pawel Danielewicz and Paul Martin is shown in Fig.~\ref{fig:pngf2}.

The success story of nonequilibrium Green functions and the tremendous impact of Keldysh's paper \cite{keldysh64} is clearly reflected in the next meetings of the conference series and their proceedings \cite{bonitz_pngf3,bonitz_jpcs_10,vanleeuwen_jpcs_13,verdozzi_jpcs16} culminating in the present issue of the proceedings of the 2018 conference.

\subsection{Current research fields based on NEGF}\label{ss:negf-current}
During the last three decades NEGF have seen a dramatic increase in attention. This is mostly due to the increase in computing capabilities that have made direct solutions of the Keldysh-Kadanoff-Baym equations possible. Applications have been developed for a large number of fields where many-body effects, correlations and non-quasiparticle behavior are of relevance. This includes transport in metals \cite{RammerRMP},
semiconductor optics and transport \cite{haug_2008_quantum}, nanostructures \cite{balzer_prb_9}, atoms and molecules \cite{dahlen_prl_07,balzer_pra_10_2}, plasma physics \cite{bonitz_99_cpp,kremp_99_pre}, nuclear matter \cite{schenke_06,koehler_06}, cosmology \cite{garny_2009,kainulainen_2010}, transport properties of strongly correlated cold fermionic atoms \cite{schluenzen_prb16,schluenzen_cpp16}, among others.

\section{Other research topics of L.V. Keldysh}\label{s:other}

\subsection{Strong Field Ionization}\label{ss:strong-field}
Cited more than 5500
times, Keldysh's paper \cite{keldysh64_field} presented the first quantum theory of the ionization of an atom by an intense laser field. The paper introduced optical tunneling, multi-photon ionization, and above-threshold ionization, experimentally observed about 15
years later, e.g. \cite{chin_jpb_85}. Keldysh presented the first nonlinear quantum-mechanical calculation of the ionization probability of an atom in a strong electromagnetic field. Starting from the time-dependent bound state wave function (we follow the notation of Ref.~\cite{ReviewStrongFieldIonization}) $\Psi_0(t)=\psi_0(\textbf{r})e^{iI_pt/\hbar}$ he computes the transition probability amplitude of the electron into a time-dependent continuum state in the presence of the field (i.e. a Volkov state \cite{volkov_34}),
\begin{equation}
    M(\textbf{p}) = -\frac{i}{\hbar}\int_{-\infty}^{\infty}dt\,\langle \Psi_\textbf{p}(t)|V_{\rm int}(t)|\Psi_0(t)\rangle,
    \label{eq:ioniz-prob}
\end{equation}
where $I_p$ denotes the ionization potential, $\textbf{p}$ the canonical momentum, and $V_{\rm int}$ the interaction potential of the electron with the field. Note that the explicit forms of $\Psi_\textbf{p}$ and $V_{\rm int}$ depend on the chosen gauge, so the analysis requires some care. Indeed, many suggested modifications or improvements  of Keldysh's work led to gauge-dependent results giving rise to debates in the community, for details see \cite{ReviewStrongFieldIonization}.  The momentum distribution of the photoelectrons is then $dW(\textbf{p})= |M(\textbf{p})^2|d^3p$, and the total ionization probability is the momentum integral of $dW$. Using the dipole approximation for the field and neglecting Coulomb interaction and relativistic effects on $\Psi_\textbf{p}$ Keldysh was able to obtain closed expressions for the ionization probability.
The result contains an important dimensionless parameter -- the ``Keldysh parameter'',
\begin{equation}
    \gamma = \sqrt{2mI_p}\,\frac{\omega}{eE_0}\,,
    \label{eq:keldysh-parameter}
\end{equation}
which determines the
boundary between multiphoton and tunneling regimes.
Here $\omega$ and $E_0$ are, respectively, the frequency and amplitude of the exciting electric field. The Keldysh parameter describes the ratio of the characteristic momentum of the electron in the bound state, $\sqrt{2mI_p}$, to the momentum the electron gains from the field, $eE_0/\omega$.
For $\gamma < 1$ ($\gamma > 1$), ionization is dominated by the tunnel (multiphoton) mechanism. In case of a monochromatic field and $\gamma>1$, multiphoton absorption is possible if the atom absorbs at least
\begin{equation}
    N_{\rm min} = \frac{I_p+U_p}{\hbar \omega} + 1
    \nonumber
\end{equation}
photons which includes the average kinetic energy of the free electron in the field (``ponderomotive potential''), $U_p=(eE_0)^2/(4m\omega^2)$. If the photon number exceeds $N_{\rm min}$, ionization will lead to distinct peaks in the photoelectron energy spectrum -- which has been called ``above threshold ionization'' -- and has been accurately verified experimentally. With the dramatic progress in laser technology and the availability of coherent radiation sources from  the infrared range to x-rays, these effects have achieved fundamental importance in countless fields.

Keldysh's theory triggered a tremendous wave of further improvements of the theory that include Coulomb interaction, relativistic effects or the field-induced modification of the bound states (Stark effect).
The analysis of ionization processes was extended to more complex atoms, molecules and semiconductors, and similar approaches were developed for relativistic effects such as pair creation (Schwinger mechanism). For additional information and references, the reader is referred to the review \cite{ReviewStrongFieldIonization}.

\subsection{Excitons and electron-hole systems}\label{ss:eh}
Keldysh made important contributions to semiconductor physics. He was early on interested in the many--exciton problem in semiconductors. In his work with Yu.V.
Kopaev \cite{keldysh_ptt_64} he introduced the new concept of an excitonic insulator. Actually this was a new mechanism of a metal--insulator transition. In later works by Keldysh and
his collaborators it was shown conclusively, that there are
no superfluidity properties in this model \cite{keldysh_jetp_72}, as was initially
suspected by some authors, and he moved to the study nonequilibrium systems of excitons, appearing
under intense laser pumping of semiconductors, where
superfluidity of excitons was shown to be possible \cite{keldysh_jetp_68}\footnote{Here also a less known paper of 1972 on the coherent state of excitons should be mentioned, that was recently reprinted in Ref.~\cite{ISI:000424395100006} with comments by M. Sadovski. There Keldysh derived what is now called the Gross-Pitaevkii equation
for coherent excitonic state in an external electromagnetic field.}.
However at that time (1968) Keldysh realized, that in
most semiconductors (with multiple bands) the nonequilibrium system of many excitons actually transforms
into an electron--hole quantum liquid (where excitons are ionized), forming electron--hole droplets. Interestingly
enough, this idea was expressed only in his summary
talk at the Moscow International Conference on Semiconductors \cite{keldysh_poluprov68} and was not published anywhere for a rather long
time. However, it immediately stimulated  experimental studies, and electron--hole droplets were soon discovered, leading to many further experimental and theoretical works on this new state of matter. Essentially, he supervised these works around the Soviet Union, continuing to introduce new concepts, such as the phonon ``wind'' in the system of electron--hole droplets \cite{keldysh_jetp_76}. An overview of the field of electron-hole droplets can be found in the review \cite{tihkodeev_ufn_85}. Electron-hole droplet formation was also verified in \textit{ab initio} quantum Monte Carlo simulations \cite{bonitz_jpa_03}. The problem of limited life time of electron-hole pairs in optically excited semiconductors can be overcome with indirect excitons predicted by Lozovik and co-workers \cite{lozovik75} which have interesting superfluidity properties \cite{boening_prb_11}.
\subsection{Further research results}\label{ss:other}
Even though Keldysh is mainly famous for real-time Green functions, strong-field ionization and the theory of excitons, he has made important contributions to many other fields.
\subsubsection{Franz-Keldysh effect}\label{sss:fke}
 It was a natural question to ask whether the Franz-Keldysh effect (the shift of the absorption edge due to an applied static electric field) could be extended to a situation where the absorbing sample was placed in a time-dependent field.  Indeed, early theoretical work addressed some aspects of this situation\cite{Yacoby,Rebane}. Experimentally, however, sufficiently strong time-dependent fields were not available until the first free-electron lasers started operation. A detailed study was published in Ref.~\cite{DFKE}, where the excitonic absorption of a quantum well system was studied as a function of the frequency of the impinging strong THz field emanating from the Santa Barbara free-electron laser. The theory developed for this situation agreed very well with the observations. The theory combines three concepts in whose development the pioneering ideas by Keldysh were crucial: strong field effects in semiconductors, excitonic dynamics, and nonequilibrium Green's functions.  It is remarkable that all three ingredients originate from the same author.

\subsubsection{Transport in mesoscopic systems}
The scattering theory of transport, developed by Landauer and B{\"u}ttiker \cite{Landauer,Buttiker}, which expresses the conductance of a mesoscopic sample in terms of its transmission properties, is -- despite of its huge success and importance -- only valid for systems where electron-electron or electron-phonon interactions can be ignored.  The Keldysh diagram technique, which allows for a systematic treatment of interactions, is particularly well-suited for deriving extensions of the Landauer-B{\"u}ttiker formalism. The Keldysh technique, as applied to transport physics, was introduced in the Western literature in an important series of papers by Caroli, Combescot, and co-workers \cite{Caroli1,Caroli2,Combescot,Caroli4}. These papers were mainly concerned with tunneling through a single barrier (including interactions with localized states and phonons in the barrier), but a real breakthrough occurred in 1992, when Meir and Wingreen showed \cite{MW} that the calculation of the conductance through a quantum dot with arbitrary interactions could be formulated in a similar manner. Literally thousands of papers have examined transport in situations where interactions are important. 

One example is the tunneling of electrons between a tip and a metal through a single adsorbed molecule (or atom) in a scanning tunneling microscope. The Keldysh technique provides an elegant way to describe the inelastic stationary electron tunneling  with emission and absorption of vibrational excitations of the molecule, interactions
with the phonon baths in the substrate and tip, as well as the overheating of the molecule and its resulting motion -- hopping or rotation
\cite{Tikhodeev2001,Mii2003,Arseyev2006,Tikhodeev2009,Shchadilova2013}.
Keldysh's theory provides the theoretical basis for inelastic tunneling electron spectroscopy, single molecule
chemistry and motors, for details see, e.g., the text book \cite{Ueba2011}.

The approach can be generalized to time-dependent situations \cite{JWM}, or situations where the partitioning of the system into separate leads and a central region must be re-examined \cite{Stefanucci}. The next level of abstraction can be achieved by formulating the nonequilibrium theory in a field-theory language.  This powerful formulation has found a very large number of applications, which are reviewed, e.g. in a recent advanced text-book \cite{Kamenev}.  The field-theory formulation honors Keldysh by employing many technical terms that commemorate their inventor, e.g., Keldysh rotation, or Keldysh action.
\subsubsection{The Rytova-Keldysh potential}
In 1979 Kel\-dysh considered the Coulomb interaction in thin semiconductor and semimetal films, and proposed a form for the interaction potential between charged particles in such systems \cite{keldysh_1979_JETPL}. (Work along similar lines was reported earlier by Rytova \cite{rytova}). A central theme in condensed matter physics in our millenium is concerned with two-dimensional materials, such as graphene, or transition metal dichalcogenides.  The Rytova-Keldysh potential forms an important ingredient in the physics of these materials.  Recent developments are reviewed, e.g. in~\cite{Kezerashvili2016} where many references to related work can be found.

\subsubsection{Stochastic methods applied to the Keldysh contour}\label{sss:dmc}
The idea of treating quantum many-body systems out of equilibrium on the Keldysh time contour has been extended to various other methods. A stochastic sampling method of Feynman diagrams was developed by Werner \textit{et al.} and is known as diagrammatic Monte Carlo, see \cite{diag_mc_werner_09} and references therein.
Diagrammatic Monte Carlo extends earlier equilibrium simulations such as the continuous-time quantum Monte Carlo method for fermions \cite{ctmc_rubtsov_05} to arbitrary nonequilibrium situations. While it formally can treat strongly coupled systems and is successfully used in condensed matter systems and for cold atoms, it suffers from the dynamic fermion sign problem that strongly limits the simulation duration.


\section{The Keldysh school}\label{s:school}

Actually Keldysh's scientific interests were much broader
than one could judge from his list of publications.  This is, in part, reflected in the broad range of topics his PhD students worked on, see Sec.~\ref{s:school}.
One of us (MS) recalls ``I first met him in 1969 when I was a third year student of the Ural State University and attended his lectures on exciton condensation and electron--hole droplets at the  famous winter school on theoretical physics ``Kourovka'' near Sverdlovsk (now Ekaterinburg). In 1971
I became his PhD student at the Lebedev Institute in Moscow and, to my surprise, he proposed to me a PhD topic related to the construction of the theory of ``liquid semiconductors''---a research field developed previously in the experimental works of the Ioffe--Regel group in Leningrad and still lacking serious
theoretical foundation. This reflected Keldysh's interest in the general theory of electrons in disordered systems, being only developed at that time in the classical
works of Neville Mott, Ilya Lifshits and Philip Anderson. In the following years we tried (in fact more or less in vain!) to construct such a theory. Our main idea was to produce a theoretical model of the pseudogap---a concept introduced
by Mott on qualitative grounds to explain electronic properties of amorphous and liquid semiconductors. Here we were successful and formulated an exactly solvable model of the pseudogap, based on the summation of a complete series of Feynman diagrams
for a simplified 1D model. Actually Keldysh declined co--autorship, so these results appeared under my name only, forming the ground for my future work in many years to follow, leading eventually  to the studies of the pseudogap problem in high-$T_C$ superconductors. This model was, in fact, a generalization of a similar diagram
summation in Keldysh's studies of doped semiconductors, which appeared only in his dissertation (1965) and was later used or rediscovered by others. These are only few of many examples of his unpublished results. Most of them he was writing in large notebooks at his home, which some of his students were lucky enough to see.''

\begin{table*}[h]
\centering
  \caption{List of Keldysh's PhD (above the line) and master students (below) in chronological order, their year of graduation and their scientific topics. See also the list of references at the end of the paper.}
\begin{tabular}{|l|c|l|}
\hline
Name & Graduation & Research topics \\
\hline
Yu.~V.~Kopaev     & 1965 &  Semimetal-dielectric phase transitions\\
D.~I.~Khomskii     & 1969 & Systems with strong electronic correlations \\
R.~R.~Guseinov     & 1971  & Electron-phonon interaction in systems with excitonic instabilities \\
M.~V.~Sadovskii    & 1974 & Liquid semiconductors, Pseudogap,
Disorder and Fluctuation effects on the 1D Peierls transition\\
A.~P.~Silin        & 1975 & Condensation of excitons in semiconductors\\
B.~A.~Volkov       & 1976 & Electronic properties of semiconductors with structural instabilities \\
A.~V.~Vinogradov   & 1976 & Electronic mechanisms of light absorption of dielectrics in transparency range\\
E.~A.~Andryushin   & 1977 & Electron-hole liquid in layered semiconductors\\
V.~S.~Babichenko   & 1977  & Electron-hole liquid in strongly anisotropic semiconductors and semimetals\\
T.~A.~Onishchenko  & 1977 & Electron-hole liquid in strong magnetic field\\
V.~E.~Bisti        & 1978 & Exciton interactions in semiconductors\\
S.~G.~Tikhodeev    & 1980 & Interaction of electron-hole liquid in semiconductors with deformations. 
\\
&& Nonequilibrium diagram technique for relaxation processes\\
A.~L.~Ivanov & 1983 & Intensive electromagnetic wave in a direct-gap semiconductor \\
I.~M.~Sokolov & 1984 & Localization in the Anderson model with correlated site energies, percolation theory \\
P.~I.~Arseev        & 1986 & Electrodynamics of rough surfaces of metals and semiconductors\\
N.~S.~Maslova      & 1987 & Resonant interaction of light with a system of nonlinear oscillators.\\
&& Non-equilibrium transport through correlated systems\\
N.~A.~Gippius & 1988 & 
Quantum reflection of an exciton from the surface of an electron-hole droplet.\\
&& Interaction of electromagnetic radiation with semiconductors\\
\hline
S.~S.~Fanchenko & 1975  & Generalized diagram technique of non-equilibrium processes. \\
&& The problem of arbitrary initial conditions \\
 \hline
\end{tabular}
  \label{tab:students}
\end{table*}

The list of Keldysh's PhD students includes Yu. V. Kopaev, D.I. Khomski, R.R. Guseinov, V.S. Babichenko, B.A. Volkov, M.V. Sadovskii, A.P. Silin,
V.E. Bisti, A.V. Vinogradov, S.G. Tikhodeev, E.A. Andryushin, T.A. Onishchenko, N.S. Maslova, and P.I. Arseev,
and their year of graduation and research topics
are presented in table \ref{tab:students}.
Many of them became successful scientists themselves. Kopaev, Khomskii, Volkov, Sadovskii, Tikhodeev, Ivanov, Arseev, Maslova, and Gippius later did their habilitation. Gippius, Khomskii, Sadovskii, Tikhodeev, and Vinogradov became professors. Arseev became a corresponding member and Kopaev and Sadovskii full members of the Russian Academy of Sciences.


\section{Conclusions}\label{s:conclusion}
There have been a number of obituaries for Keldysh in the U.S. \cite{keldysh_physics-today17} and in Russia  \cite{ufn_keldysh_2017} that have covered various sides of Keldysh's scientific work and personality. The 2017 special issue of Physics Uspekhi (issue 11, volume 60) covers in detail Keldysh's scientific work. There is no need to reproduce this material here. Instead, we have taken the particular angle of view on Keldysh that concentrates on his contributions to nonequilibrium many-body physics, in general, and nonequilibrium Green functions, in particular. Keldysh's single paper on the subject \cite{keldysh64} has dramatically changed the whole field providing us and future generations with a strict mathematical basis and an extremely powerful and fully general tool -- nonequilbrium diagram technique. This has allowed the NEGF approach to being introduced in an enormously broad range of areas of physics and quantum chemistry, not just as a tool for recovering familiar kinetic equations or deriving improved approximations but, more and more, as a practical tool for quantitative analysis of time-dependent processes on all time scales. Judging by the impressive increase in the number of publications on NEGF, Keldysh's work has provided a tremendously fertile ground for theory developments in many decades to come.

\begin{acknowledgement}
We are grateful to World Scientific Publishing for the permission to reprint Keldysh's article from the PNGF II proceedings \cite{keldysh_pngf2} as a supplement to this paper and to D. Semkat for providing the LaTeX source. We thank J.-P. Joost for technical assistence with the formatting of this article. APJ is supported by the Danish National Research Foundation, Project DNRF103.
\end{acknowledgement}

%

\begin{thebibliography}{[100]}

\bibitem{ISI:A1958WT97700025}
 \textsc{L.~Keldysh},
{Behavior of Non-Metallic Crystals in Strong Electric Fields},
 \jr{{Soviet Physics JETP-USSR}} \textbf{{6}}({4}), {763--770} ({1958}).


\bibitem{ISI:A1958WX85500028}
 \textsc{L.~Keldysh},
{Influence of the Lattice Vibrations of a Crystal on the Production of
  Electron-Hole Pairs in a Strong Electrical Field},
 \jr{{Soviet Physics JETP-USSR}} \textbf{{7}}({4}), {665--668} ({1958}).


\bibitem{KeldyshEffect}
 \textsc{L.~Keldysh},
{The Effect of a Strong Electric Field on the Optical Properties of Insulating
  Crystals},
 \jr{{Soviet Physics JETP-USSR}} \textbf{{7}}({5}), {788--790} ({1958}).


\bibitem{ISI:A1960WE64800018}
 \textsc{B.~Vul},  \textsc{E.~Zavaritskaia},  and  \textsc{L.~Keldysh},
{Impurity Conductivity of Germanium at Low Temperatures},
 \jr{{Doklady Akademii Nauk SSSR}} \textbf{{135}}({6}), {1361--1363} ({1960}).


\bibitem{ISI:A1960WN97500020}
 \textsc{L.~Keldysh},
{Kinetic Theory of Impact Ionization in Semiconductors},
 \jr{{Soviet Physics JETP-USSR}} \textbf{{10}}({3}), {509--518} ({1960}).


\bibitem{ISI:A1963WM20800048}
 \textsc{L.~Keldysh},
{Optical Characteristics of Electrons with a Band Energy Spectrum in a Strong
  Electric Field},
 \jr{{Soviet Physics JETP-USSR}} \textbf{{16}}({2}), {471--474} ({1963}).


\bibitem{keldysh_ptt_62}
 \textsc{L.~Keldysh},
{Effect of Ultrasonics on the Electron Spectrum of Crystals},
 \jr{{Soviet Physics-Solid State}} \textbf{{4}}({8}), {1658--1659} ({1963}).


\bibitem{ISI:A1963WP01800029}
 \textsc{L.~Keldysh} and  \textsc{Y.~Kopaev},
{The Energy Spectrum of a Degenerate Semiconductor with an Ionic Lattice},
 \jr{{Soviet Physics-Solid State}} \textbf{{5}}({5}), {1026--1030} ({1963}).


\bibitem{keldysh_jetp_64}
 \textsc{L.~Keldysh},
{Deep Levels in Semiconductors},
 \jr{{Soviet Physics JETP-USSR}} \textbf{{18}}({1}), {253--260} ({1964}).


\bibitem{ISI:A1964WT61000005}
 \textsc{L.~Keldysh} and  \textsc{G.~Proshko},
{Infrared Absorption in Highly Doped Germanium},
 \jr{{Soviet Physics-Solid State}} \textbf{{5}}({12}), {2481--2488} ({1964}).


\bibitem{ISI:A1964WT61500020}
 \textsc{V.~Bagaev},  \textsc{Y.~Berozashvili},  \textsc{B.~Vul},
  \textsc{E.~Zavaritskaya},  \textsc{L.~Keldysh},  and  \textsc{A.~Shotov},
{Concerning the Energy Level Spectrum of Heavily Doped Gallium Arsenide},
 \jr{{Soviet Physics-Solid State}} \textbf{{6}}({5}), {1093--1098} ({1964}).


\bibitem{keldysh64}
 \textsc{L.~Keldysh},
{Diagram Technique for Nonequilibrium Processes},
 \jr{{Soviet Physics JETP-USSR}} \textbf{{20}}({4}), {1018} ({1965}),
[Zh.~Eksp.~Teor.~Fiz.~\textbf{47}, 1515 (1964)].


\bibitem{keldysh64_field}
 \textsc{L.~Keldysh},
{Ionization in Field of a Strong Electromagnetic Wave},
 \jr{{Soviet Physics JETP-USSR}} \textbf{{20}}({5}), {1307} ({1965}),
{[ZhETF \textbf{47}, 1945--1957 (1964)]}.


\bibitem{ISI:A19657071700023}
 \textsc{L.~Keldysh},
{Concerning Theory of Impact Ionization in Semiconductors},
 \jr{{Soviet Physics JETP-USSR}} \textbf{{21}}({6}), {1135} ({1965}).


\bibitem{keldysh_ptt_64}
 \textsc{L.~Keldysh} and  \textsc{Y.~Kopaev},
{Possible Instability of Semimetallic State Toward Coulomb Interaction},
 \jr{{Soviet Physics Solid State, USSR}} \textbf{{6}}({9}), {2219} ({1965}),
[Fiz. Tverd. Tela~\textbf{6}, 2791 (1964)].


\bibitem{ISI:A19657168300010}
 \textsc{L.~Keldysh},
{Superconductivity In Nonmetallic Systems},
 \jr{{Soviet Physics Uspekhi-USSR}} \textbf{{8}}({3}), {496} ({1965}).


\bibitem{ISI:A19668668700009}
 \textsc{V.~Bagaev},  \textsc{Y.~Berozashvili},  and  \textsc{L.~Keldysh},
{Electrooptical Effect in GaAs},
 \jr{{JETP Letters-USSR}} \textbf{{4}}({9}), {246} ({1966}).


\bibitem{ISI:A19667931900016}
 \textsc{L.~Keldysh} and  \textsc{T.~Tratas},
{Dynamic Narrowing of Paramagnetic Resonance Lines in a Compensated
  Semiconductor},
 \jr{{Soviet Physics Solid State, USSR}} \textbf{{8}}({1}), {64} ({1966}).


\bibitem{ISI:A19679362600012}
 \textsc{L.~Keldysh} and  \textsc{A.~Kozlov},
{Collective Properties of Large-Radius Excitons},
 \jr{{JETP Letters-USSR}} \textbf{{5}}({7}), {190} ({1967}).


\bibitem{keldysh_jetp_68}
 \textsc{L.~Keldysh} and  \textsc{A.~Kozlov},
{Collective Properties of Excitons in Semiconductors},
 \jr{{Soviet Physics JETP-USSR}} \textbf{{27}}({3}), {521} ({1968}),
[Zh.~Eksp.~Teor.~Fiz.~\textbf{54}, 978 (1968)].


\bibitem{ISI:A1969D031000004}
 \textsc{V.~Bagaev},  \textsc{Y.~Berozashvili},  and  \textsc{L.\,a. Keldysh},
{Anisotropy of Polarized-Light Absorption Produced in GaAs and CdTe Crystals by
  a Strong Electric Field},
 \jr{{JETP Letters-USSR}} \textbf{{9}}({3}), {108} ({1969}).


\bibitem{ISI:A1969E591300010}
 \textsc{L.~Keldysh} and  \textsc{M.~Pkhakadze},
{Conductivity of Semiconductors Under Pinch-Effect Conditions},
 \jr{{JETP Letters-USSR}} \textbf{{10}}({6}), {169} ({1969}).


\bibitem{ISI:A1969E811600006}
 \textsc{V.~Bagaev},  \textsc{T.~Galkina},  \textsc{O.~Gogolin},  and
  \textsc{L.~Keldysh},
{Motion of Electron-Hole Drops in Germanium},
 \jr{{JETP Letters-USSR}} \textbf{{10}}({7}), {195} ({1969}).


\bibitem{ISI:A1970F063000017}
 \textsc{L.~Keldysh},  \textsc{O.~Konstantinov},  and  \textsc{V.~Perel},
{Polarization Effects in Interband Absorption of Light in Semiconductors
  Subjected to a Strong Electric Field},
 \jr{{Soviet Physics Semiconductors-USSR}} \textbf{{3}}({7}), {876} ({1970}).


\bibitem{ISI:A1970I354600011}
 \textsc{L.~Keldysh},
{Electron-Hole Drops in Semiconductors},
 \jr{{Soviet Physics Uspekhi-USSR}} \textbf{{13}}({2}), {292} ({1970}).


\bibitem{ISI:A1972M332000004}
 \textsc{B.~Kadomtsev},  \textsc{R.~Sagdeev},  \textsc{L.~Keldysh},  and
  \textsc{I.~Kobzarev},
{On A.A. TYAPKIN's article ``Expression of General Properties of Physical
  Processes in Space-and-Time Metric of Special Theory of Relativity''},
 \jr{{Uspekhi Fizicheskikh Nauk}} \textbf{{106}}({4}), {660} ({1972}).


\bibitem{keldysh_jetp_72}
 \textsc{R.~Guseinov} and  \textsc{L.~Keldysh},
{Nature of Phase-Transitions under Excitonic Instability Conditions of a
  Crystal Electron Spectrum},
 \jr{{Zhurnal Eksperimentalnoi i Teoreticheskoi Fiziki}} \textbf{{63}}({12}),
  {2255--2263} ({1972}).


\bibitem{ISI:A1973P786000039}
 \textsc{L.~Keldysh} and  \textsc{A.~Silin},
{Electron-Hole Liquids in Semiconductors in Magnetic-Field},
 \jr{{FIZIKA TVERDOGO TELA}} \textbf{{15}}({5}), {1532--1535} ({1973}).


\bibitem{ISI:A1974T394700028}
 \textsc{L.~Keldysh},  \textsc{A.~Manenkov},  \textsc{V.~Milyaev},  and
  \textsc{G.~Mikhailova},
{Microwave Breakdown and Exciton Condensation in Germanium},
 \jr{{Zhurnal Eksperimentalnoi i Teoreticheskoi Fiziki}} \textbf{{66}}({6}),
  {2178--2190} ({1974}).


\bibitem{ISI:A1975AU26400005}
 \textsc{L.~Keldysh} and  \textsc{S.~Tikhodeev},
{Absorption of Ultrasound by Electron-Hole Drops in a Semiconductor},
 \jr{{JETP Letters}} \textbf{{21}}({10}), {273--274} ({1975}).


\bibitem{ISI:A1975AT03500030}
 \textsc{L.~Keldysh} and  \textsc{A.~Silin},
{Electron-Hole Fluid in Polar Semiconductors},
 \jr{{Zhurnal Eksperimentalnoi i Teoreticheskoi Fiziki}} \textbf{{69}}({3}),
  {1053--1057} ({1975}).


\bibitem{ISI:A1976CE06600002}
 \textsc{L.~Keldysh},
{Phonon Wind and Dimensions of Electron-Hole Drops in Semiconductors},
 \jr{{JETP Letters}} \textbf{{23}}({2}), {86--89} ({1976}).


\bibitem{ISI:A1976DB30000006}
 \textsc{L.~Keldysh} and  \textsc{T.~Onishchenko},
{Electron Liquid in a Superstrong Magnetic-Field},
 \jr{{JETP Letters}} \textbf{{24}}({2}), {59--62} ({1976}).


\bibitem{ISI:A1976DJ19200006}
 \textsc{E.~Andryushin},  \textsc{V.~Babichenko},  \textsc{L.~Keldysh},
  \textsc{T.~Onishchenko},  and  \textsc{A.~Silin},
{Electron-Hole Liquid in Strongly Anisotropic Semiconductors and Semimetals},
 \jr{{JETP Letters}} \textbf{{24}}({4}), {185--189} ({1976}).


\bibitem{keldysh_jetp_76}
 \textsc{V.~Bagaev},  \textsc{L.~Keldysh},  \textsc{N.~Sibeldin},  and
  \textsc{V.~Tsvetkov},
{Phonon Wind Drag of Excitons and Electron-Hole Drops},
 \jr{{Zhurnal Eksperimentalnoi i Teoreticheskoi Fiziki}} \textbf{{70}}({2}),
  {702--716} ({1976}).


\bibitem{ISI:A1976BR33200035}
 \textsc{V.~Bagaev},  \textsc{N.~Zamkovets},  \textsc{L.~Keldysh},
  \textsc{N.~Sibeldin},  and  \textsc{V.~Tsvetkov},
{Kinetics of Exciton Condensation in Germanium},
 \jr{{Zhurnal Eksperimentalnoi i Teoreticheskoi Fiziki}} \textbf{{70}}({4}),
  {1500--1521} ({1976}).


\bibitem{ISI:A1977CT30500023}
 \textsc{L.~Keldysh} and  \textsc{S.~Tikhodeev},
{Ultrasound Absorption by Electron-Hole Drops in Semiconductor},
 \jr{{Fizika Tverdogo Tela}} \textbf{{19}}({1}), {111--117} ({1977}).


\bibitem{ISI:A1977DV91300039}
 \textsc{E.~Andrushin},  \textsc{L.~Keldysh},  and  \textsc{A.~Silin},
{Electron-Hole Liquid and Metal-Dielectric Phase-Transition in Layer Systems},
 \jr{{Zhurnal Eksperimentalnoi i Teoreticheskoi Fiziki}} \textbf{{73}}({3}),
  {1163--1173} ({1977}).


\bibitem{ISI:A1978FR24900010}
 \textsc{L.~Keldysh},
{Metal-Dielectric Transformation Under Light Action},
 \jr{{Vestnik Moskovskovo Universiteta Seria 3 Fizika Astronomia}}
  \textbf{{19}}({4}), {86--90} ({1978}).


\bibitem{ISI:A1979JN00600015}
 \textsc{L.~Keldysh},
{Coulomb Interaction in Thin Semiconductor and Semimetal Films},
 \jr{{JETP Letters}} \textbf{{29}}({11}), {658--661} ({1979}).


\bibitem{keldysh_1979_JETPL}
 \textsc{L.~Keldysh},
{Polaritons in Thin Semiconducting-Films},
 \jr{{JETP Letters}} \textbf{{30}}({4}), {224--227} ({1979}),
{[ZHETF Letters \textbf{29}, 658 (1979)]}.


\bibitem{ISI:A1980LJ42200006}
 \textsc{V.~Bagaev},  \textsc{M.~Bonchosmolovskii},  \textsc{T.~Galkina},
  \textsc{L.~Keldysh},  and  \textsc{A.~Poyarkov},
{Entrainment of Electron-Hole Drops by a Strain Pulse Produced as a Result of
  Laser Irradiation of Germanium},
 \jr{{JETP Letters}} \textbf{{32}}({5}), {332--335} ({1980}).


\bibitem{ISI:A1980KN51300038}
 \textsc{E.~Andriushyn},  \textsc{L.~Keldysh},  \textsc{V.~Sanina},  and
  \textsc{A.~Silin},
{Electron-Hole Liquid in Thin Semiconducting-Films},
 \jr{{Zhurnal Eksperimentalnoi i Teoreticheskoi Fiziki}} \textbf{{79}}({4}),
  {1509--1517} ({1980}).


\bibitem{ISI:A1981MB17700017}
 \textsc{L.~Keldysh} and  \textsc{A.~Kechek},
{On the Dielectric-Constant of The Non-Polar Fluid},
 \jr{{Doklady Akademii Nauk SSSR}} \textbf{{259}}({3}), {575--578} ({1981}).


\bibitem{ISI:A1982PJ79700021}
 \textsc{A.~Ivanov} and  \textsc{L.~Keldysh},
{The Propagation of Powerful Electromagnetic-Waves in Semiconductors under the
  Resonant Excitation of Excitons},
 \jr{{Doklady Akademii Nauk SSSR}} \textbf{{264}}({6}), {1363--1366} ({1982}).


\bibitem{ISI:A1983QB18700036}
 \textsc{A.~Ivanov} and  \textsc{L.~Keldysh},
{Modification of the Polariton and Phonon-Spectra of a Semiconductor in the
  Presence of an Intense Electromagnetic-Wave},
 \jr{{Zhurnal Eksperimentalnoi i Teoreticheskoi Fiziki}} \textbf{{84}}({1}),
  {404--421} ({1983}).


\bibitem{ISI:A1984AFY8900005}
 \textsc{N.~Gippius},  \textsc{V.~Zavaritskaya},  \textsc{L.~Keldysh},
  \textsc{V.~Milyaev},  and  \textsc{S.~Tikhodeev},
{Quantum Nature Of the Reflection of an Exciton from the Surface of an
  Electron-Hole Drop},
 \jr{{JETP Letters}} \textbf{{40}}({10}), {1235--1238} ({1984}).


\bibitem{ISI:A1984TJ11900023}
 \textsc{P.~Elyutin},  \textsc{L.~Keldysh},  and  \textsc{A.~Kechek},
{The Resonance Dielectric Permittivity of Nonpolar Liquids},
 \jr{{Optika i Spektroskopia}} \textbf{{57}}({2}), {282--287} ({1984}).


\bibitem{ISI:A1986C643700031}
 \textsc{L.~Keldysh} and  \textsc{S.~Tikhodeev},
{High-Intensity Polariton Wave Near the Stimulated Scattering Threshold},
 \jr{{Zhurnal Eksperimentalnoi i Teoreticheskoi Fiziki}} \textbf{{90}}({5}),
  {1852--1870} ({1986}).


\bibitem{ISI:A1986D472300009}
 \textsc{L.~Keldysh} and  \textsc{S.~Tikhodeev},
{Nonstationary Mandelstam-Brillouin Scattering of an Intense Polariton Wave},
 \jr{{Zhurnal Eksperimentalnoi i Teoreticheskoi Fiziki}} \textbf{{91}}({1}),
  {78--85} ({1986}).


\bibitem{ISI:A1986F487200028}
 \textsc{N.~Gippius},  \textsc{L.~Keldysh},  and  \textsc{S.~Tikhodeev},
{Mandelstam-Brilloiun Scattering of an Incoherent Polariton Wave},
 \jr{{Zhurnal Eksperimentalnoi i Teoreticheskoi Fiziki}} \textbf{{91}}({6}),
  {2263--2275} ({1986}).


\bibitem{ISI:A1988P896100047}
 \textsc{L.~Keldysh},
{Excitons and Polaritons in Semiconductor Insulator Quantum Wells and
  Superlattices},
 \jr{{Superlattices and Microstructures}} \textbf{{4}}({4-5}), {637--642}
  ({1988}).


\bibitem{ISI:A1991FE66200024}
 \textsc{A.~Ivanov},  \textsc{L.~Keldysh},  and  \textsc{V.~Panashchenko},
{Low-Threshold Exciton-Biexciton Optical Stark-Effect in Direct-Gap
  Semiconductors},
 \jr{{Zhurnal Eksperimentalnoi i Teoreticheskoi Fiziki}} \textbf{{99}}({2}),
  {641--658} ({1991}).


\bibitem{ISI:A1992JQ13000097}
 \textsc{A.~Ivanov},  \textsc{L.~Keldysh},  and  \textsc{V.~Panashchenko},
{Nonlinear Optical-Response of Interacting Excitons},
 \jr{{Institute of Physics Conference Series}}({126}), {431--436} ({1992}).


\bibitem{ISI:A1992JR28400011}
 \textsc{L.~Keldysh},
{Excitonic Molecules in Nonlinear Optical-Response},
 \jr{{Physica Status Solidi B}} \textbf{{173}}({1}), {119--128} ({1992}).


\bibitem{ISI:A1992JT05900011}
 \textsc{L.~Keldysh},
{Coherent Excitonic Molecules},
 \jr{{Solid State Communications}} \textbf{{84}}({1-2}), {37--43} ({1992}).


\bibitem{ISI:A1993ML21500095}
 \textsc{N.~Gippius},  \textsc{T.~Ishihara},  \textsc{L.~Keldysh},
  \textsc{E.~Muljarov},  and  \textsc{S.~Tikhodeev},
{Dielectrically Confined Excitons and Polaritons in Natural Superlattices -
  Perovskite Lead Iodide Semiconductors},
 \jr{{Journal de Physique IV}} \textbf{{3}}({C5}), {437--440} ({1993}),
{3rd International Conference on Optics of Excitons in Confined Systems, Univ
  Montpellier II, Montpellier, France, Aug 30-Sep 02, 1993}.


\bibitem{ISI:A1994QA79500021}
 \textsc{N.~Gippius},  \textsc{S.~Tikhodeev},  and  \textsc{L.~Keldysh},
{Polaritons in Semiconductor-Insulator Superlattices with Nonlocal Excitonic
  Response},
 \jr{{Superlattics and Microstructures}} \textbf{{15}}({4}), {479--482}
  ({1994}).


\bibitem{ISI:A1995QQ83700001}
 \textsc{L.~Keldysh},
{Correlations in the Coherent Transient Electron-Hole System},
 \jr{{Physica Status Solidi B}} \textbf{{188}}({1}), {11--27} ({1995}),
{4th International Workshop on Nonlinear Optics and Excitation Kinetics in
  Semiconductors (NOEKS IV), Gosen, Germany, Nov 06-10, 1994}.


\bibitem{ISI:A1997XR96400071}
 \textsc{A.~Ivanov},  \textsc{H.~Wang},  \textsc{J.~Shah},  \textsc{T.~Damen},
  \textsc{L.~Keldysh},  \textsc{H.~Haug},  and  \textsc{L.~Pfeiffer},
{Coherent transient in photoluminescence of excitonic molecules in GaAs quantum
  wells},
 \jr{{Physical Review B}} \textbf{{56}}({7}), {3941--3951} ({1997}).


\bibitem{ISI:000071319400004}
 \textsc{L.~Keldysh},
{Excitons in semiconductor-dielectric nanostructures},
 \jr{{Physica Status Solidi A}} \textbf{{164}}({1}), {3--12} ({1997}),
{5th International Meeting on Optics of Excitons in Confined Systems (OECS 5),
  G\"ottingen, Germany, Aug 10-14, 1997}.


\bibitem{ISI:000072424800001}
 \textsc{A.~Ivanov},  \textsc{H.~Haug},  and  \textsc{L.~Keldysh},
{Optics of excitonic molecules in semiconductors and semiconductor
  microstructures},
 \jr{{Physics Reports}} \textbf{{296}}({5-6}), {237--336} ({1998}).


\bibitem{ISI:000087302200002}
 \textsc{Q.~Vu},  \textsc{H.~Hang},  and  \textsc{L.~Keldysh},
{Dynamics of the electron-hole correlation in femtosecond pulse excited
  semiconductors},
 \jr{{Solid State Communications}} \textbf{{115}}({2}), {63--65} ({2000}).


\bibitem{ISI:000179600900006}
 \textsc{L.~Keldysh},
{Biexcitons at high densities},
 \jr{{Pysica Status Solidi B}} \textbf{{234}}({1}), {17--22} ({2002}).


\bibitem{ISI:000184336600038}
 \textsc{F.~Klappenberger},  \textsc{K.~Renk},  \textsc{R.~Summer},
  \textsc{L.~Keldysh},  \textsc{B.~Rieder},  and  \textsc{W.~Wegscheider},
{Electric-field-induced reversible avalanche breakdown in a GaAs microcrystal
  due to cross band gap impact ionization},
 \jr{{Applied Physics Letters}} \textbf{{83}}({4}), {704--706} ({2003}).


\othercit
\bibitem{keldysh_pngf2}
 \textsc{L.~Keldysh},
{Real-Time Nonequilbirium Green's Functions},
 in: Progress in Nonequilibrium Green's functions II, edited by M.~Bonitz and
  D.~Semkat,  (World Scientific Publ., Singapore, 2003),  pp.\,4--17,
[reprinted as supplemental material].


\bibitem{ISI:000225020200039}
 \textsc{J.~Reithmaier},  \textsc{G.~Sek},  \textsc{A.~Loffler},
  \textsc{C.~Hofmann},  \textsc{S.~Kuhn},  \textsc{S.~Reitzenstein},
  \textsc{L.~Keldysh},  \textsc{V.~Kulakovskii},  \textsc{T.~Reinecke},  and
  \textsc{A.~Forchel},
{Strong coupling in a single quantum dot-semiconductor microcavity system},
 \jr{{Nature}} \textbf{{432}}({7014}), {197--200} ({2004}).


\bibitem{ISI:000230494900007}
 \textsc{N.~Gippius},  \textsc{S.~Tikhodeev},  \textsc{L.~Keldysh},  and
  \textsc{V.~Kulakovskii},
{Hard excitation of stimulated polariton-polariton scattering in semiconductor
  microcavities},
 \jr{{Physics Uspekhi}} \textbf{{48}}({3}), {306--312} ({2005}).


\bibitem{ISI:000230238900002}
 \textsc{Y.~Osipov},  \textsc{V.~Sadovnichii},  \textsc{V.~Kozlov},
  \textsc{O.~Krokhin},  \textsc{N.~Zefirov},  \textsc{E.~Velikhov},
  \textsc{G.~Dobrovol'skii},  \textsc{L.~Keldysh},  \textsc{S.~Nikol'skii},
  \textsc{Y.~Tret'yakov},  \textsc{K.~Frolov},  \textsc{V.~Khain},
  \textsc{E.~Chazov},  \textsc{V.~Yanin},  \textsc{V.~Kabanov},
  \textsc{A.~Solzhenitsyn},  \textsc{L.~Faddeev},  \textsc{A.~Andreev},
  \textsc{G.~Chernyi},  \textsc{V.~Lunin},  \textsc{G.~Dobrovol'skii},
  \textsc{D.~Pushcharovskii},  \textsc{V.~Stepin},  \textsc{A.~Derevyanko},
  \textsc{A.~Kudelin},  \textsc{R.~Nigmatulin},  \textsc{T.~Oizerman},
  \textsc{N.~Dikanskii},  \textsc{N.~Plate},  \textsc{V.~Kostyuk},  and
  \textsc{V.~Urusov},
{Joint scientific session of the General Meeting of the Russian Academy of
  Sciences and the Academic Council of Moscow State University named after M.V.
  Lomonosov, dedicated to the 250th anniversary of Moscow State University},
 \jr{{Herald of the Russian Academy of Sciences}} \textbf{{75}}({3}),
  {214--270} ({2005}).


\bibitem{ISI:000237842200120}
 \textsc{G.~Sek},  \textsc{C.~Hofmann},  \textsc{J.~Reithmaier},
  \textsc{A.~Loffler},  \textsc{S.~Reitzenstein},  \textsc{M.~Kamp},
  \textsc{L.~Keldysh},  \textsc{V.~Kulakovskii},  \textsc{T.~Reinecke},  and
  \textsc{A.~Forchel},
{Investigation of strong coupling between single quantum dot excitons and
  single photons in pillar microcavities},
 \jr{{Physica E}} \textbf{{32}}({1-2}), {471--475} ({2006}),
{12th International Conference on Modulated Semiconductor Structures (MSS12),
  Albuquerque, NM, JUL 10-15, 2005}.


\bibitem{ISI:000237728200050}
 \textsc{S.~Reitzenstein},  \textsc{A.~Loffler},  \textsc{C.~Hofmann},
  \textsc{A.~Kubanek},  \textsc{M.~Kamp},  \textsc{J.~Reithmaier},
  \textsc{A.~Forchel},  \textsc{V.~Kulakovskii},  \textsc{L.~Keldysh},
  \textsc{I.~Ponomarev},  and  \textsc{T.~Reinecke},
{Coherent photonic coupling of semiconductor quantum dots},
 \jr{{Optics Letters}} \textbf{{31}}({11}), {1738--1740} ({2006}).


\bibitem{ISI:000239932300003}
 \textsc{S.~Reitzenstein},  \textsc{C.~Hofmann},  \textsc{A.~Loeffler},
  \textsc{A.~Kubanek},  \textsc{J.\,P. Reithmaier},  \textsc{M.~Kamp},
  \textsc{V.\,D. Kulakovskii},  \textsc{L.\,V. Keldysh},  \textsc{T.\,L.
  Reinecke},  and  \textsc{A.~Forchel},
{Strong and weak coupling of single quantum dot excitons in pillar
  microcavities},
 \jr{{Physica Status Solidi B}} \textbf{{243}}({10}), {2224--2228} ({2006}),
{8th International Workshop on Nonlinear Optics and Excitation Kinetics In
  Semiconductors (NOEKS 8), M\"unster, Germany, FEB 20-24, 2006}.


\bibitem{ISI:000243731300005}
 \textsc{L.\,V. Keldysh},  \textsc{V.\,D. Kulakovskii},
  \textsc{S.~Reitzenstein},  \textsc{M.\,N. Makhonin},  and
  \textsc{A.~Forchel},
{Interference effects in the emission spectra of quantum dots in high-quality
  cavities},
 \jr{{JETP Letters}} \textbf{{84}}({9}), {494--499} ({2006}).


\bibitem{ISI:000242102500035}
 \textsc{S.~Reitzenstein},  \textsc{A.~Loffler},  \textsc{A.~Kubanek},
  \textsc{C.~Hofmann},  \textsc{M.~Kamp},  \textsc{J.\,P. Reithmaier},
  \textsc{A.~Forchel},  \textsc{V.\,D. Kulakovskii},  \textsc{L.\,V. Keldysh},
  \textsc{I.\,V. Ponomarev},  and  \textsc{T.\,L. Reinecke},
{Coherent photonic coupling of semiconductor quantum dots (vol 31, pg 1738,
  2006)},
 \jr{{Optics Letters}} \textbf{{31}}({23}), {3507} ({2006}).


\bibitem{ISI:000392143400001}
 \textsc{L.\,V. Keldysh},
{Dynamic Tunneling},
 \jr{{Herald of the Russian Academy of Sciences}} \textbf{{86}}({6}),
  {413--425} ({2016}).


\bibitem{ISI:000424395100006}
 \textsc{L.\,V. Keldysh},
{Coherent states of excitons},
 \jr{{Physics Uspekhi}} \textbf{{60}}({11}), {1180--1186} ({2017}).


\bibitem{ISI:000424395100007}
 \textsc{L.\,V. Keldysh},
{Multiphoton ionization by a very short pulse},
 \jr{{Physics Uspekhi}} \textbf{{60}}({11}), {1187--1193} ({2017}).


\bibitem{ISI:A1992KH81500002}
 \textsc{A.~Aleksandrov},  \textsc{Z.~Alverov},  \textsc{N.~Basov},
  \textsc{E.~Velikhov},  \textsc{A.~Gonchar},  \textsc{A.~Dynkin},
  \textsc{L.~Keldysh},  \textsc{D.~Knorre},  \textsc{V.~Kotelnikov},
  \textsc{G.~Mesyats},  \textsc{Y.~Osipov},  \textsc{V.~Pokrovskii},
  \textsc{B.~Saltykov},  \textsc{V.~Subbotin},  \textsc{L.~Faddeev},
  \textsc{E.~Chelyshev},  and  \textsc{V.~Shorin},
{State Research Centers (Discussion in The Russian-Academy-of-Sciences)},
 \jr{{Vestnik Rossiskoi Akademii Nauk}}({12}), {14--29} ({1992}).


\bibitem{ISI:A1992JG61900007}
 \textsc{L.~Keldysh},
{Russian Science at The Approaching Market},
 \jr{{Vestnik Rossiskoi Akademii Nauk}}({3}), {45--52} ({1992}).


\bibitem{ISI:A1992HD70800001}
 \textsc{Z.~Alferov},  \textsc{V.~Ginzburg},  \textsc{V.~Goldanskii},
  \textsc{L.~Keldysh},  \textsc{V.~Maslov},  \textsc{A.~Spirin},  and
  \textsc{V.~Keilisborok},
{Urgent Appeal for Help},
 \jr{{Chemical \& Engineering News}} \textbf{{70}}({7}), {2} ({1992}).


\bibitem{ISI:A1996TX10100004}
 \textsc{A.~Gonchar},  \textsc{A.~Spirin},  \textsc{Y.~Osipov},
  \textsc{D.~Knoppe},  \textsc{N.~Shilo},  \textsc{L.~Faddeev},
  \textsc{V.~Sadovnichii},  \textsc{Z.~Alferov},  \textsc{E.~Velikhov},
  \textsc{V.~Subbotin},  \textsc{V.~Martynov},  \textsc{V.~Kudryavtsev},
  \textsc{I.~Makarov},  \textsc{E.~Chelyshev},  \textsc{A.~Prokhorov},
  \textsc{M.~Styrikovich},  \textsc{V.~Orel},  \textsc{V.~Sokolov},
  \textsc{P.~Simonov},  \textsc{L.~Keldysh},  \textsc{G.~Semin},
  \textsc{A.~Egorov},  \textsc{B.~Saltykov},  \textsc{N.~Basov},  and
  \textsc{N.~Laverov},
{What doctrine of science advancement is needed by Russia? Discussion in the
  RAS Presidium},
 \jr{{Vestnik Rossiskoi Akademii Nauk}} \textbf{{66}}({1}), {16--25} ({1996}).


\bibitem{ISI:A1996UX63100002}
 \textsc{Z.~Alferov},  \textsc{Y.~Osipov},  \textsc{A.~Spirin},
  \textsc{V.~Subbotin},  \textsc{E.~Velikhov},  \textsc{N.~Laverov},
  \textsc{G.~Golitsyn},  \textsc{A.~Gonchar},  \textsc{I.~Makarov},
  \textsc{A.~Prokhorov},  \textsc{V.~Sobolev},  and  \textsc{L.~Keldysh},
{The Ioffe Physico-Technical Institute in the new economic conditions -
  Discussion in the RAS Presidium},
 \jr{{Vestnik Rossiskoi Akademii Nauk}} \textbf{{66}}({6}), {491--498}
  ({1996}).


\bibitem{ISI:A1997XF16000005}
 \textsc{V.~Ginzburg} and  \textsc{L.~Keldysh},
{The age qualification in elections to the Academy of Sciences cannot be
  tolerated},
 \jr{{Vestnik Rossiskoi Akademii Nauk}} \textbf{{67}}({4}), {321--322}
  ({1997}).


\bibitem{ISI:000085196500001}
 \textsc{A.~Boyarchuk} and  \textsc{L.~Keldysh},
{From a physics laboratory to the General Physics and Astronomy Division},
 \jr{{Uspekhi Fizicheskikh Nauk}} \textbf{{169}}({12}), {1289--1298} ({1999}).


\bibitem{keldysh_physics-today17}
 \textsc{F.~Capasso},  \textsc{P.~Corkum},  \textsc{O.~Kocharovskaya},
  \textsc{L.~Pitaevskii},  and  \textsc{M.~Sadovskii},
{Leonid Keldysh},
 \jr{Physics Today} \textbf{70}, 75 (2017),
shortened version of the present text.


\bibitem{FranzEffect}
 \textsc{W.~Franz},
{Einfluss eines elektrischen Feldes aurf eine optische Absorptionskante},
 \jr{Z. Naturforsch. Teil A} \textbf{{\bf 13}}, 484 (1958).


\bibitem{ReviewStrongFieldIonization}
 \textsc{S.\,V. Popruzhenko},
{Keldysh theory of strong field ionization: history, applications, difficulties
  and perspectives},
 \jr{J. Phys. B: At. Mol. Opt. Phys.} \textbf{{\bf 47}}, 204001 (2014).


\othercit
\bibitem{bonitz_qkt}
 \textsc{M.~Bonitz},
Quantum Kinetic Theory, 2 edition, Teubner-Texte zur Physik (Springer, 2016).


\bibitem{martin-schwinger_59}
 \textsc{P.\,C. Martin} and  \textsc{J.~Schwinger},
Theory of many-particle systems. i,
 \jr{Phys. Rev.} \textbf{115}(Sep), 1342--1373 (1959).


\othercit
\bibitem{kadanoff-baym-book}
 \textsc{L.~Kadanoff} and  \textsc{G.~Baym},
Quantum Statistical Mechanics (Benjamin, New York, 1962).


\othercit
\bibitem{martin_pngf1}
 \textsc{P.~Martin},
Quantum kinetic equations,
 in: Progress in Nonequilibrium Green's functions, edited by M.~Bonitz,  (World
  Scientific Publ., Singapore, 2000),  pp.\,2--16.


\othercit
\bibitem{baym_pngf1}
 \textsc{G.~Baym},
Conservation laws and the quantum theory of transport: The early days,
 in: Progress in Nonequilibrium Green's functions, edited by M.~Bonitz,  (World
  Scientific Publ., Singapore, 2000),  pp.\,17--32.


\othercit
\bibitem{abrikosov_pngf2}
 \textsc{A.~Abrikosov},
Story about the temperature technique,
 in: Progress in Nonequilibrium Green's functions II, edited by M.~Bonitz and
  D.~Semkat,  (World Scientific Publ., Singapore, 2003),  pp.\,2--3.


\othercit
\bibitem{supplement}
{Supplementary material, include URL}.


\bibitem{DANIELEWICZ_84_ap2}
 \textsc{P.~Danielewicz},
{Quantum theory of nonequilibrium processes II. Application to nuclear
  collisions},
 \jr{Annals of Physics} \textbf{152}(2), 305 -- 326 (1984).


\bibitem{bonitz_pngf3}
 \textsc{M.~Bonitz} and  \textsc{A.~Filinov},
{Progress in Nonequilibrium Green's Functions III},
 \jr{Journal of Physics: Conference Series} \textbf{35}(1) (2006).


\othercit
\bibitem{schaefer_wegener}
 \textsc{W.~Sch\"afer} and  \textsc{M.~Wegener},
Semiconductor Optics and Transport Phenomena (Springer, 2002).


\bibitem{bornath_cpp17}
 \textsc{T.~Bornath},  \textsc{W.~Kraeft},  \textsc{R.~Redmer},
  \textsc{G.~R\"opke},  \textsc{M.~Schlanges},  \textsc{W.~Ebeling},  and
  \textsc{M.~Bonitz},
{A tribute to Dietrich Kremp},
 \jr{Contributions to Plasma Physics} \textbf{57}(10), 434--440 (2017).


\othercit
\bibitem{kremp_schlanges_kraeft_book}
 \textsc{D.~Kremp},  \textsc{M.~Schlanges},  and  \textsc{W.~Kraeft},
Quantum Statistics of Nonideal Plasmas (Springer, 2005).


\othercit
\bibitem{haug_2008_quantum}
 \textsc{H.~Haug} and  \textsc{A.\,P. Jauho},
Quantum Kinetics in Transport and Optics of Semiconductors (Springer, 2008).


\othercit
\bibitem{pngf2}
 \textsc{M.~Bonitz} and  \textsc{D.~Semkat},
{Progress in Nonequilibrium Green's Functions II}, Progress in Nonequilibrium
  Green's Functions (World Scientific, 2003).


\bibitem{bonitz_jpcs_10}
 \textsc{M.~Bonitz} and  \textsc{K.~Balzer},
{Progress in Nonequilibrium Green's Functions IV},
 \jr{Journal of Physics: Conference Series} \textbf{220}(1), 011001 (2010).


\bibitem{vanleeuwen_jpcs_13}
 \textsc{R.~van Leeuwen},  \textsc{R.~Tuovinen},  and  \textsc{M.~Bonitz},
{Progress in Nonequilibrium Green's Functions V (PNGF V)},
 \jr{Journal of Physics: Conference Series} \textbf{427}(1), 011001 (2013).


\bibitem{verdozzi_jpcs16}
 \textsc{C.~Verdozzi},  \textsc{A.~Wacker},  \textsc{C.\,O. Almbladh},  and
  \textsc{M.~Bonitz},
{Progress in Non-equilibrium Green's Functions (PNGF VI)},
 \jr{Journal of Physics: Conference Series} \textbf{696}(1), 011001 (2016).


\bibitem{RammerRMP}
 \textsc{J.~Rammer} and  \textsc{H.~Smith},
{Quantum field-theoretical methods in transport theory of metals},
 \jr{Reviews in Modern Physics} \textbf{{\bf 58}}, 323--359 (1986).


\bibitem{balzer_prb_9}
 \textsc{K.~Balzer},  \textsc{M.~Bonitz},  \textsc{R.~van Leeuwen},
  \textsc{A.~Stan},  and  \textsc{N.\,E. Dahlen},
{Nonequilibrium Green's function approach to strongly correlated few-electron
  quantum dots},
 \jr{Phys. Rev. B} \textbf{79}(Jun), 245306 (2009).


\bibitem{dahlen_prl_07}
 \textsc{N.\,E. Dahlen} and  \textsc{R.~van Leeuwen},
{Solving the Kadanoff-Baym Equations for Inhomogeneous Systems: Application to
  Atoms and Molecules},
 \jr{Phys. Rev. Lett.} \textbf{98}(Apr), 153004 (2007).


\bibitem{balzer_pra_10_2}
 \textsc{K.~Balzer},  \textsc{S.~Bauch},  and  \textsc{M.~Bonitz},
Time-dependent second-order {Born} calculations for model atoms and molecules
  in strong laser fields,
 \jr{Phys. Rev. A} \textbf{82}, 033427 (2010).


\bibitem{bonitz_99_cpp}
 \textsc{M.~Bonitz},  \textsc{T.~Bornath},  \textsc{D.~Kremp},
  \textsc{M.~Schlanges},  and  \textsc{W.\,D. Kraeft},
Quantum kinetic theory for laser plasmas. dynamical screening in strong fields,
 \jr{Contrib. Plasma Phys.} \textbf{{\bf 39}}(4), 329--347 (1999).


\bibitem{kremp_99_pre}
 \textsc{D.~Kremp},  \textsc{T.~Bornath},  \textsc{M.~Bonitz},  and
  \textsc{M.~Schlanges},
Quantum kinetic theory of plasmas in strong laser fields,
 \jr{Phys. Rev. E} \textbf{{\bf 60}}(Oct), 4725--4732 (1999).


\bibitem{schenke_06}
 \textsc{B.~Schenke} and  \textsc{C.~Greiner},
Nonequilibrium description of dilepton production in heavy ion reactions,
 \jr{Journal of Physics: Conference Series} \textbf{35}(1), 398 (2006).


\bibitem{koehler_06}
 \textsc{H.\,S. K\"ohler},
{Beyond the quasi-particle picture in nuclear matter calculations using Green's
  function techniques},
 \jr{Journal of Physics: Conference Series} \textbf{35}(1), 384 (2006).


\bibitem{garny_2009}
 \textsc{M.~Garny} and  \textsc{M.\,M. M\"uller},
{Kadanoff-Baym equations with non-Gaussian initial conditions: The equilibrium
  limit},
 \jr{Phys. Rev. D} \textbf{80}(Oct), 085011 (2009).


\bibitem{kainulainen_2010}
 \textsc{M.~Herranen},  \textsc{K.~Kainulainen},  and  \textsc{P.\,M. Rahkila},
Coherent quasiparticle approximation (cqpa) and nonlocal coherence,
 \jr{Journal of Physics: Conference Series} \textbf{220}(1), 012007 (2010).


\bibitem{schluenzen_prb16}
 \textsc{N.~Schl\"unzen},  \textsc{S.~Hermanns},  \textsc{M.~Bonitz},  and
  \textsc{C.~Verdozzi},
Dynamics of strongly correlated fermions: \textit{Ab initio} results for two
  and three dimensions,
 \jr{Phys. Rev. B} \textbf{93}(Jan), 035107 (2016).


\bibitem{schluenzen_cpp16}
 \textsc{N.~Schl\"unzen} and  \textsc{M.~Bonitz},
{Nonequilibrium Green Functions Approach to Strongly Correlated Fermions in
  Lattice Systems},
 \jr{Contributions to Plasma Physics} \textbf{56}(1), 5--91 (2016).


\bibitem{chin_jpb_85}
 \textsc{S.\,L. Chin},  \textsc{F.~Yergeau},  and  \textsc{P.~Lavigne},
Tunnel ionisation of xe in an ultra-intense co 2 laser field (10 14 w cm -2 )
  with multiple charge creation,
 \jr{Journal of Physics B: Atomic and Molecular Physics} \textbf{18}(8), L213
  (1985).


\bibitem{volkov_34}
 \textsc{D.~Volkov},
 \jr{Z. Phys.} \textbf{94}, 250 (1934).


\othercit
\bibitem{keldysh_poluprov68}
 \textsc{L.~Keldysh},
{Concluding remarks},
 in: {Proceedings of the IXth International Conference of Semiconductor
  Physics, Moscow 23-29 July 1968},  (Nauka, Leningrad, 1969),
  pp.\,1303--1312.


\bibitem{tihkodeev_ufn_85}
 \textsc{S.\,G. Tikhodeev},
The electron-hole liquid in a semiconductor,
 \jr{Soviet Physics Uspekhi} \textbf{28}(1), 1 (1985).


\bibitem{bonitz_jpa_03}
 \textsc{M.~Bonitz},  \textsc{D.~Semkat},  \textsc{A.~Filinov},
  \textsc{V.~Golubnychyi},  \textsc{D.~Kremp},  \textsc{D.\,O. Gericke},
  \textsc{M.\,S. Murillo},  \textsc{V.~Filinov},  \textsc{V.~Fortov},
  \textsc{W.~Hoyer},  and  \textsc{S.\,W. Koch},
Theory and simulation of strong correlations in quantum {C}oulomb systems,
 \jr{J. Phys. A: Math. Gen.} \textbf{{\bf 36}}(22), 5921 (2003).


\bibitem{lozovik75}
 \textsc{Y.\,E. Lozovik} and  \textsc{V.\,I. Yudson},
{Feasibility of superfluidity of paired spatially separated electrons and holes
  - new superconductivity mechansim},
 \jr{JETP Letters} \textbf{{\bf 22}}, 274--276 (1975).


\bibitem{boening_prb_11}
 \textsc{J.~B\"oning},  \textsc{A.~Filinov},  and  \textsc{M.~Bonitz},
Crystallization of an exciton superfluid,
 \jr{Phys. Rev. B} \textbf{84}, 075130 (2011).


\bibitem{Yacoby}
 \textsc{Y.~Yacoby},
{High-frequency Franz-Keldysh effect},
 \jr{Phys. Rev.} \textbf{{\bf 169}}, 610 (1968).


\bibitem{Rebane}
 \textsc{Y.\,T. Rebane},
{Semiconductor and dielectric energy-band reconstruction in a field of
  intensive weak self-absorbing light waves},
 \jr{Fiz. Tverd. Tela} \textbf{{\bf 27}}, 1364 (1985).


\bibitem{DFKE}
 \textsc{K.\,B. Nordstrom},  \textsc{K.~Johnsen},  \textsc{S.\,J. Allen},
  \textsc{A.\,P. Jauho},  \textsc{B.~Birnir},  \textsc{J.~Kono},
  \textsc{T.~Noda},  \textsc{H.~Akiyama},  and  \textsc{H.~Sakaki},
{Excitonic dynamical Franz-Keldysh effect},
 \jr{Phys. Rev. Lett.} \textbf{{\bf 81}}, 457--460 (1998).


\bibitem{Landauer}
 \textsc{R.~Landauer},
{Spatial varaion of currents and fields due to localized scatterers in metallic
  conduction},
 \jr{IBM Journal of research and development} \textbf{{\bf 1}}, 223--231
  (1957).


\bibitem{Buttiker}
 \textsc{M.~B{\"u}ttiker},
{4-terminal phase-coherent conductance},
 \jr{Phys. Rev. Lett.} \textbf{{\bf 57}}, 1761--1764 (1986).


\bibitem{Caroli1}
 \textsc{C.~Caroli},  \textsc{R.~Combescot},  \textsc{P.~Nozieres},  and
  \textsc{D.~Saint-James},
{Direct calculation of tunneling current},
 \jr{J. Phys. C Sol. St. Physics} \textbf{{\bf 4}}, 916 (1971).


\bibitem{Caroli2}
 \textsc{C.~Caroli},  \textsc{R.~Combescot},  \textsc{D.~Lederer},
  \textsc{P.~Nozieres},  and  \textsc{D.~Saint-James},
{Direct calculation of tunneling current. 2. Free electron description},
 \jr{J. Phys. C Sol. St. Physics} \textbf{{\bf 4}}, 2598 (1971).


\bibitem{Combescot}
 \textsc{R.~Combescot},
{Direct calculation of tunneling current. 3. Effect of localized impurity
  states in barrier},
 \jr{J. Phys. C Sol. St. Physics} \textbf{{\bf 4}}, 2611 (1971).


\bibitem{Caroli4}
 \textsc{C.~Caroli},  \textsc{D.\,S. James},  \textsc{R.~Combescot},  and
  \textsc{P.~Nozieres},
{Direct calculation of tunneling current. 4. Electron-phonon interaction
  effects},
 \jr{J. Phys. C Sol. St. Physics} \textbf{{\bf 5}}, 21 (1971).


\bibitem{MW}
 \textsc{Y.~Meir} and  \textsc{N.\,S. Wingreen},
{Landauer formula for the current through an interacting electron region},
 \jr{Phys. Rev. Lett} \textbf{{\bf 68}}, 2512--2515 (1992).


\bibitem{Tikhodeev2001}
 \textsc{S.~Tikhodeev},  \textsc{M.~Natario},  \textsc{K.~Makoshi},
  \textsc{T.~Mii},  and  \textsc{H.~Ueba},
{Contribution to a theory of vibrational scanning tunneling spectroscopy of
  adsorbates. Nonequilibrium Green's function approach},
 \jr{Surf. Sci.} \textbf{493}(1-3), 63--70 (2001).


\bibitem{Mii2003}
 \textsc{T.~Mii},  \textsc{S.\,G. Tikhodeev},  and  \textsc{H.~Ueba},
Spectral features of inelastic electron transport via a localized state,
 \jr{Phys. Rev. B} \textbf{68}(20), 205406 (2003).


\bibitem{Arseyev2006}
 \textsc{P.\,I. Arseyev} and  \textsc{N.\,S. Maslova},
Tunneling current induced phonon generation in nanostructures,
 \jr{Pis'ma v ZhETF} \textbf{84}(2), 99--104 (2006),
[JETP Letters, 2006, Vol. \textbf{84}, No. 2, pp. 93--98.].


\bibitem{Tikhodeev2009}
 \textsc{S.\,G. Tikhodeev} and  \textsc{H.~Ueba},
{How Vibrationally Assisted Tunneling with STM Affects the Motions and
  Reactions of Single Adsorbates},
 \jr{Phys. Rev. Lett.} \textbf{102}(24), 246101 (2009).


\bibitem{Shchadilova2013}
 \textsc{Y.\,E. Shchadilova},  \textsc{S.\,G. Tikhodeev},
  \textsc{M.~Paulsson},  and  \textsc{H.~Ueba},
{Rotation of a Single Acetylene Molecule on Cu(001) by Tunneling Electrons in
  STM},
 \jr{Phys. Rev. Lett.} \textbf{111}(Oct), 186102 (2013).


\othercit
\bibitem{Ueba2011}
 \textsc{H.~Ueba},  \textsc{S.\,G. Tikhodeev},  and  \textsc{B.\,N.\,J.
  Persson},
{in: Current-Driven Phenomena in Nanoelectronics},
 (Pan Stanford Publishing Pte. Ltd., 2011), chap.~2,  pp.\,26--89,
Theory of inelastic tunneling current-driven motions of single adsorbates.


\bibitem{JWM}
 \textsc{A.\,P. Jauho},  \textsc{N.\,S. Wingreen},  and  \textsc{Y.~Meir},
{Time-dependent transport in interacting and noninteracting resonant-tunneling
  systems},
 \jr{Physical Review B} \textbf{{\bf 50}}, 5528--5544 (1994).


\bibitem{Stefanucci}
 \textsc{G.~Stefanucci} and  \textsc{C.\,O. Almbladh},
{Time-dependent partition-free approach in resonant tunneling systems},
 \jr{Phys. Rev. B} \textbf{{\bf 69}}, 195318 (2004).


\othercit
\bibitem{Kamenev}
 \textsc{A.~Kamenev},
Field Theory of Non-Equilibrium Systems (Cambridge University Press, 2011).


\bibitem{rytova}
 \textsc{N.\,S. Rytova},
{Coulomb interaction of electrons in a thin film},
 \jr{Doklady Akademii Nauk SSSR} \textbf{{\bf 163}}, 1118 (1965).


\bibitem{Kezerashvili2016}
 \textsc{R.\,Y. Kezerashvili} and  \textsc{S.\,M. Tsiklauri},
Trion and biexciton in monolayer transition metal dichalcogenides,
 \jr{Few-Body Systems} \textbf{58}(1), 18 (2016).


\bibitem{diag_mc_werner_09}
 \textsc{P.~Werner},  \textsc{T.~Oka},  and  \textsc{A.\,J. Millis},
{Diagrammatic Monte Carlo simulation of nonequilibrium systems},
 \jr{Phys. Rev. B} \textbf{79}(Jan), 035320 (2009).


\bibitem{ctmc_rubtsov_05}
 \textsc{A.\,N. Rubtsov},  \textsc{V.\,V. Savkin},  and  \textsc{A.\,I.
  Lichtenstein},
{Continuous-time quantum Monte Carlo method for fermions},
 \jr{Phys. Rev. B} \textbf{72}(Jul), 035122 (2005).


\othercit
\bibitem{ufn_keldysh_2017}
{Physics Uspekhi \textbf{60} (11), 1065 (2017)}.


\end{thebibliography}

\providecommand{\WileyBibTextsc}{}
\let\textsc\WileyBibTextsc
\providecommand{\othercit}{}
\providecommand{\jr}[1]{#1}
\providecommand{\etal}{~et~al.}

\end{document}